\documentclass[
 reprint,
 superscriptaddress,
 amsmath,amssymb,
 aps,
prl,
]{revtex4-2}

\usepackage{graphicx}
\usepackage{dcolumn}
\usepackage{bm}
\usepackage{siunitx}
\usepackage{xspace}
\usepackage{xcolor}
\DeclareSIUnit{\gauss}{G}

\newcommand{\ket}[1]{\ensuremath{\lvert #1 \rangle}\xspace}%
\newcommand{\avg}[1]{\ensuremath{\langle #1 \rangle}\xspace}%
\usepackage[colorlinks=true, citecolor=blue, urlcolor=blue, linkcolor=blue]{hyperref}
\begin{document}

\newcommand{\partitle}[1]{\subsection*{#1}}

\title{Velocity-Enabled Quantum Computing with Neutral Atoms}

\author{Ohad Lib}
\email{ohad.lib@mpq.mpg.de}
   \affiliation{Max-Planck-Institut f\"{u}r Quantenoptik, 85748 Garching, Germany}
   \affiliation{Munich Center for Quantum Science and Technology (MCQST), 80799 Munich, Germany}
   
\author{Hendrik Timme}
   \affiliation{Max-Planck-Institut f\"{u}r Quantenoptik, 85748 Garching, Germany}
   \affiliation{Munich Center for Quantum Science and Technology (MCQST), 80799 Munich, Germany}

\author{Maximilian Ammenwerth}
  \affiliation{Max-Planck-Institut f\"{u}r Quantenoptik, 85748 Garching, Germany}
   \affiliation{Munich Center for Quantum Science and Technology (MCQST), 80799 Munich, Germany}
   
\author{Flavien Gyger}
   \affiliation{Max-Planck-Institut f\"{u}r Quantenoptik, 85748 Garching, Germany}
   \affiliation{Munich Center for Quantum Science and Technology (MCQST), 80799 Munich, Germany}
   
\author{Renhao Tao}
   \altaffiliation[Present address: ]{Department of Physics, Harvard University, Cambridge, Massachusetts 02138, USA}
   \affiliation{Max-Planck-Institut f\"{u}r Quantenoptik, 85748 Garching, Germany}
   \affiliation{Munich Center for Quantum Science and Technology (MCQST), 80799 Munich, Germany}
   \affiliation{Fakult\"{a}t f\"{u}r Physik, Ludwig-Maximilians-Universit\"{a}t, 80799 Munich, Germany}

\author{Shijia Sun}
   \affiliation{Max-Planck-Institut f\"{u}r Quantenoptik, 85748 Garching, Germany}
   \affiliation{Munich Center for Quantum Science and Technology (MCQST), 80799 Munich, Germany}
   \affiliation{Fakult\"{a}t f\"{u}r Physik, Ludwig-Maximilians-Universit\"{a}t, 80799 Munich, Germany}
   
\author{Immanuel Bloch}
   \affiliation{Max-Planck-Institut f\"{u}r Quantenoptik, 85748 Garching, Germany}
   \affiliation{Munich Center for Quantum Science and Technology (MCQST), 80799 Munich, Germany}
   \affiliation{Fakult\"{a}t f\"{u}r Physik, Ludwig-Maximilians-Universit\"{a}t, 80799 Munich, Germany}

\author{Johannes Zeiher}
\affiliation{Max-Planck-Institut f\"{u}r Quantenoptik, 85748 Garching, Germany}
\affiliation{Munich Center for Quantum Science and Technology (MCQST), 80799 Munich, Germany}
\affiliation{Fakult\"{a}t f\"{u}r Physik, Ludwig-Maximilians-Universit\"{a}t, 80799 Munich, Germany}

\date{\today}
\begin{abstract}
    Realizing error-corrected logical qubits is a central goal for the current development of digital quantum computers.
    Neutral atoms offer the opportunity to coherently shuttle atoms for realizing efficient quantum error correction based on long-range connectivity and parallel atom transport. 
    Nevertheless, time overheads in shuttling atoms and complex control hardware pose challenges to scaling current architectures.
    Here, we introduce atom velocity as a new degree of freedom in neutral-atom architectures tailored to quantum error correction.
    Through controlled Doppler shifts, we demonstrate velocity-selective mid-circuit state preparation and measurement on moving atoms, leaving spectator atoms unaffected. 
    Furthermore, we achieve on-the-fly local single-qubit rotations by mapping micron-scale atom displacements to the spatial phase of global control beams. 
    Complementing these techniques with CZ entangling gates with a fidelity of $99.86(4)\%$, we experimentally implement key primitives for quantum error correction and measurement-based quantum computing. We generate an eight-qubit entangled cluster state with an average stabilizer value of 0.830(4), realize an $[[4,2,2]]$ error-detection code with \SI{99.0(3)}{\percent} logical Bell-state fidelity, and perform stabilizer measurements using a flying ancilla.
    By enabling selective operations on continuously moving atoms using only global beams, this velocity-enabled architecture reduces hardware overhead while minimizing shuttling and transfer delays, opening a new pathway for fast, large-scale atom-based quantum computation.
\end{abstract}
\maketitle



Achieving fault-tolerant quantum computing requires precise control over a large number of qubits~\cite{gidney2025factor}.
Trapped neutral atoms have emerged as a promising platform in this race, with demonstrated high-fidelity operations~\cite{evered2023high,peper2025spectroscopy,muniz2025high,tsai2025benchmarking,radnaev2025universal,senoo2025high, lin2026} and scalability potential~\cite{Tao2024, pause2024supercharged,manetsch2025tweezer,lin2025ai,holman2026trapping}.
Recent demonstrations have shown the preparation, control, and continuous loading~\cite{gyger2024continuous,norcia2024iterative,chiu2025continuous,li2025fast,muniz2025repeated} of large-scale arrays, with below-threshold quantum error correction (QEC) already demonstrated~\cite{bluvstein2025fault}.

A key advantage of atomic platforms for QEC is the arbitrary connectivity enabled by coherent atom shuttling~\cite{beugnon2007two, lengwenus2010coherent,bluvstein2022quantum}, which facilitates the efficient realization of transversal logical gates and the implementation of high-rate quantum error correcting codes~\cite{gottesman2013fault,breuckmann2021quantum,bluvstein2025fault,pecorari2025high,poole2025architecture}.
Beyond long-range connectivity, atom shuttling has further been utilized for hardware-efficient qubit-selective control by moving atoms between functionally distinct spatial zones, each addressed by one or more specific laser beams~\cite{bluvstein2022quantum}.
However, such zone-based control in large-scale arrays requires atom shuttling over long distances, which comes with a significant overhead in computation time.
%
Leveraging efficient transversal operations on logical qubits~\cite{zhou2025low}, optimized hardware-aware circuit compilation~\cite{stade2024abstract,Tan2024compilingquantum}, or tightly focused laser beams for local addressing~\cite{graham2022multi, rines2025demonstration}, together with experimentally efficient QEC codes~\cite{sahay2023high} can help reduce this time overhead.
Nevertheless, scaling precise and qubit-selective control to large arrays while maintaining long-range connectivity and fast QEC cycle times remains a challenge.

\begin{figure*}[t!]
\centering
\includegraphics[width=1\textwidth]{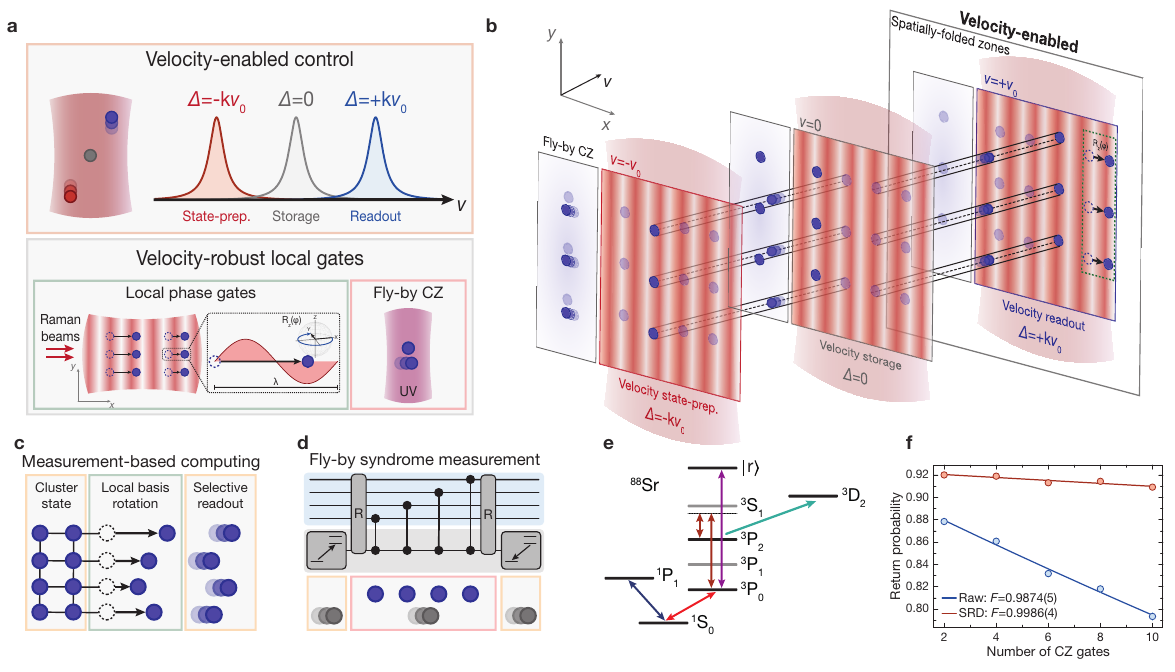}
\caption{\label{fig:1} \textbf{Velocity-enabled architecture.} \textbf{a} Building blocks for velocity-enabled control. Controlled Doppler shifts on moving atoms define velocity zones for storage, state-preparation, and measurement. Within each velocity zone, velocity-robust local single-qubit operations are performed via on-the-fly micron-scale atom displacements. Fly-by CZ gates between atoms moving at different velocities complete the set of operations that can be performed using continuously moving atoms at any velocity zone. \textbf{b} An architectural overview utilizing these building blocks, featuring velocity as a new degree of freedom. Spatially separated zones used in a standard zone-based architecture are replaced with spatially co-located velocity zones addressed by global beams with controlled detunings. On-the-fly single-qubit operations and fly-by CZ gates can be performed at any velocity zone, where spatial zones for two-qubit gates (blue shading) or single-qubit gates (red modulated zone) are still well-defined. To switch between zones, atoms are accelerated to the velocity addressed by the appropriately detuned global control beam. These concepts enable efficient realization of key primitives in measurement-based quantum computing and quantum error correction. \textbf{c} Measurement-based computation on an entangled cluster state requires local selective readout in various measurement bases. This can be achieved in our architecture through local basis rotations via atom displacement followed by velocity-selective readout on those already moving atoms. \textbf{d} For quantum error correction, syndrome measurements can be performed using a flying ancilla which is first selectively excited from the ground state to the qubit manifold, then being entangled with data qubits via fly-by CZ gates, and finally selectively measured without affecting the stationary data qubits. \textbf{e} Level structure of $^{88}\text{Sr}$ used in our experimental demonstration with the fine-structure qubit, showing the relevant transitions for state-preparation (red), measurement (blue), single-qubit control (dark red), Rydberg excitation (purple), and depumping (turquoise) used in state-resolved detection (SRD). Beam directions and further details are given in the SI~\cite{SI}. \textbf{f} Symmetric stabilizer state benchmarking (SSB)~\cite{tsai2025benchmarking} of CZ gates for static atoms with (orange) and without (blue) SRD.}
\end{figure*}

Here, we propose and realize a new architecture for neutral-atom quantum computing that enables fast on-the-fly selective operations by utilizing the atoms' velocity as a new degree of freedom (Fig.~\ref{fig:1}a, b).
When an atom moves toward a control laser beam, it experiences a Doppler-shifted frequency.
While Doppler shifts are widely used for cooling~\cite{hansch1975cooling}, they are often considered a source of errors for coherent operations.
Through controlled atom movement, we take advantage of tailored Doppler shifts and narrow-line transitions available in alkaline-earth and alkaline-earth-like atoms to perform selective mid-circuit state preparation and measurement on velocity-selected atoms. 
With sub-micron-per-microsecond-scale atomic speeds, we define spatially co-located velocity zones (Fig.~\ref{fig:1}b) where such selective operations can be realized within a few microns of continuous motion, regardless of the size of the array, and without additional local control beams.
Beyond state preparation and measurement, we further implement local single-qubit rotations via the position of moving atoms along the control beam path.
While previously explored for controlling optical qubits using displacements on the order of \SI{100}{\nano\meter}~\cite{shaw2024multi}, we demonstrate such rotations for the fine-structure (FS) qubit in $^{88}\text{Sr}$ using micron-scale moves.
Importantly, we show that such local control can be achieved by timing single-qubit gates during continuous atom movement, making it compatible with our velocity-enabled architecture.
Finally, we complement this on-the-fly single-qubit local control with fly-by CZ gates and demonstrate key primitives of measurement-based quantum computing (MBQC) and QEC (Fig.~\ref{fig:1}c, d).
%

Our experiments were performed using $^{88}\text{Sr}$-atoms trapped in a tweezer array at a wavelength of \SI{813}{nm}~\cite{tao2025universal,ammenwerth2025realization} (Fig.~\ref{fig:1}e).
In this platform, we realize a universal gate set on the fine-structure qubit, consisting of the ${^3}\text{P}_0$ and ${^3}\text{P}_2, m_J=0$ clock states, under triple-magic trapping conditions~\cite{unnikrishnan2024coherent,pucher2024fine,ammenwerth2025realization} with fast erasure imaging and state-selective detection~\cite{ammenwerth2025realization,tao2025universal} for near-ground-state-cooled atoms.
For the present work, we further improve our two-qubit gate fidelity up to \SI{99.86(4)}{\percent} with delayed erasure through state-resolved detection~\cite{tao2025universal} (Fig.~\ref{fig:1}f), and implement coherent atom shuttling in addition to a static spatial light modulator-based tweezer array using tweezers generated by a pair of crossed acousto-optic deflectors (AODs) at \SI{813}{\nano\meter}~\cite{SI}.

\begin{figure*}[t!]
\centering
\includegraphics[width=1\textwidth]{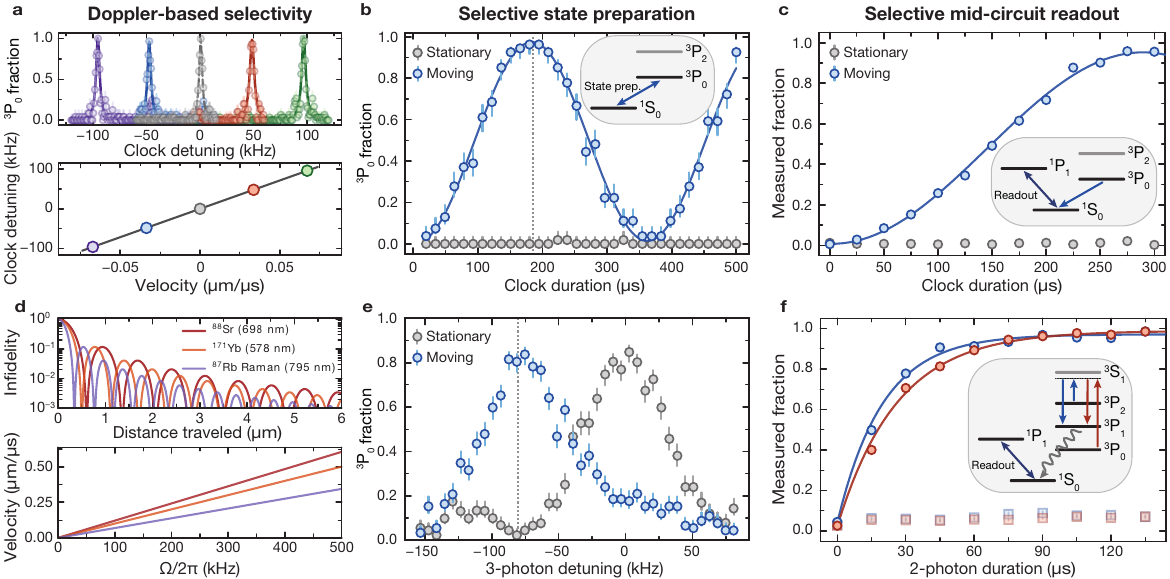}
\caption{\label{fig:2} \textbf{Velocity-selective state preparation and measurement.} \textbf{a} Atoms are moved towards or away from a \SI{698}{\nano\meter} clock laser at different velocities. The ground state to $^{3}\text{P}_{0}$ resonance is shifted due to the Doppler effect (top) with the expected linear dependence on velocity (bottom). \textbf{b} Moving selected atoms at a constant velocity of \SI{0.03}{\meter\per\second} and matching the clock-laser detuning to the Doppler-shifted atoms, Rabi oscillations are selectively performed on the moving atoms, with nearby stationary atoms remaining in the ground state. By performing a $\pi$-pulse, selective state preparation of moving atoms into the metastable qubit manifold is achieved. \textbf{c} A similar concept can be applied for velocity-selective readout, where moving atoms, initially in $^{3}\text{P}_{0}$, are transferred to the ground state. Then, they are measured via a fast mid-circuit image, leaving stationary atoms in the qubit manifold unaffected. The difference in $\pi$-times between \textbf{b} and \textbf{c} is a result of the different powers used for the \SI{698}{\nano\meter} clock laser. \textbf{d} Theoretical results for velocity selectivity on a single-qubit clock transition in strontium and ytterbium and on a Raman transition with counter-propagating beams in rubidium. An analytical curve of infidelity incurred on stationary qubits for a velocity-selective $\pi$-pulse is plotted as a function of the distance traveled by the moving atoms during the pulse (top). The infidelity is independent of Rabi frequency and decreases with the distance traveled. Zeroes corresponding to detunings where stationary atoms exhibit a $2\pi$-rotation are observed and attainable for sub-\SI{}{\micro\meter} move distances. The required velocity of the moving atoms at the first zero infidelity point is plotted as a function of Rabi frequency (bottom). \textbf{e} An order of magnitude faster state preparation and readout can be achieved via a three-photon transition, driven by 679, 688, and \SI{689}{\nano\meter} beams. A spectroscopy measurement of stationary and moving atoms is shown for a fine-tuned velocity of \SI{0.052}{\meter\per\second} in the direction of the \SI{688}{\nano\meter} beam, where stationary atoms are minimally affected. \textbf{f} A robust transfer of atoms into the ground state can also be achieved via dissipative two-photon coupling of both qubit states into $^{3}\text{P}_{1}$, where they decay and can be imaged via a fast image. Moving atoms (round points) are depumped, whereas stationary ones remain in their state (square points). This shows the potential of velocity selectivity for qubit-reset operations.}
\end{figure*}

\begin{figure*}[t!]
\centering
\includegraphics[width=1\textwidth]{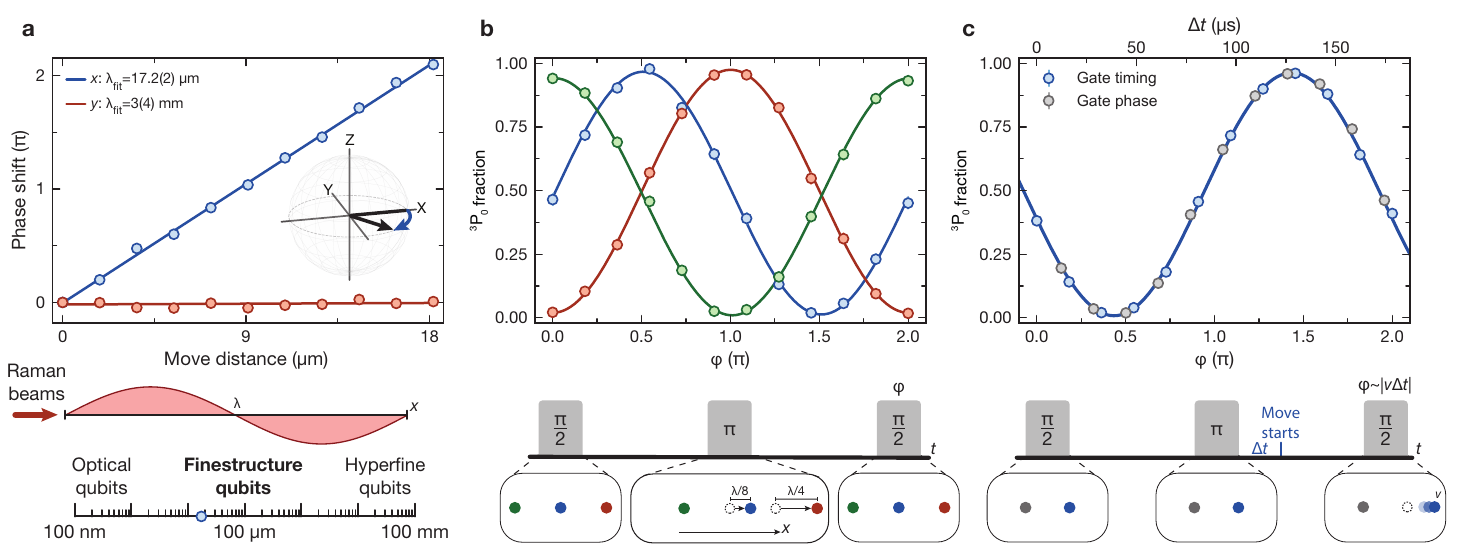}
\caption{\label{fig:3} \textbf{Fast local rotations via on-the-fly displacement.} \textbf{a} The phase acquired by a qubit displaced along the Raman beams' direction ($x$) is measured in a Ramsey sequence and shows a linear trend with the expected wavelength of the FS qubit. Moving along the perpendicular direction ($y$) shows no phase shift, as expected. The effective wavelength of the FS qubit is in an intermediate regime between optical and hyperfine qubits, allowing for convenient phase control using micron-scale displacements. \textbf{b} Local control over qubit rotation is demonstrated in an echo sequence. Different atoms are displaced by varying distances, showing the expected shift in the Ramsey signal. In the experiment, four stationary atoms were trapped in static traps, while two columns of four atoms each were trapped in movable AOD tweezers and displaced independently. \textbf{c} FS echo Ramsey measurement with static and moving atoms. Thanks to the reduced sensitivity to Doppler shifts and higher Rabi frequency of the FS qubit in comparison with optical qubits, the atoms do not need to stop during the gate operation. This is demonstrated by changing the timing of the last gate in an echo sequence with respect to the atom's movement, yielding a different phase as a function of its position at the time of the gate.}
\end{figure*}

We begin by demonstrating selective, local state preparation and detection of the metastable qubit manifold on the \SI{698}{\nano\meter} ${^1}\text{S}_0$ to ${^3}\text{P}_0$ clock transition.
We accelerate ground-state atoms trapped in movable AOD tweezers to a constant velocity along the direction of the clock beam and apply a pulse with varying detunings.
We observe a linear shift of the resonance frequency by a detuning $\Delta$ with velocity $v$ of the atoms, as expected from the Doppler effect  $\Delta = k_0v$, where $k_0$ denotes the wavenumber of the clock light (Fig.~\ref{fig:2}a).
Detuning the clock laser to the shifted resonance, we selectively drive Rabi oscillations on atoms moving at a constant velocity of $v=\SI{0.03}{\meter\per\second}$, with nearby stationary atoms remaining unaffected (Fig.~\ref{fig:2}b).
At the $\pi$-time, we achieve selective state preparation of moving atoms into the qubit manifold, with a fidelity of \SI{97(1)}{\percent}, which is equivalent to the fidelity obtained in our experiment with global control without velocity selectivity, and can be improved via erasure detection~\cite{tao2025universal}.
Already at such low velocity, the residual excitation of stationary qubits is measured to be only \SI{0.4(3)}{\percent} at the $\pi$-time, and can be reduced even further by using higher velocities.
The same concept can be applied in reverse for selective mid-circuit readout.
Atoms prepared in ${^3}\text{P}_0$ are selectively transferred to ${^1}\text{S}_0$ using a clock pulse, followed by a fast mid-circuit image of the ground state, which is currently performed with the atoms being stationary~\cite{tao2025universal}.
Varying the clock duration up to the $\pi$-time, an increasing percentage of up to \SI{96(1)}{\percent} of moving atoms are detected in the mid-circuit image, while stationary atoms remain undetected in the qubit subspace (Fig.~\ref{fig:2}c).

%
%
\begin{figure}[t]
\centering
\includegraphics[width=1\columnwidth]{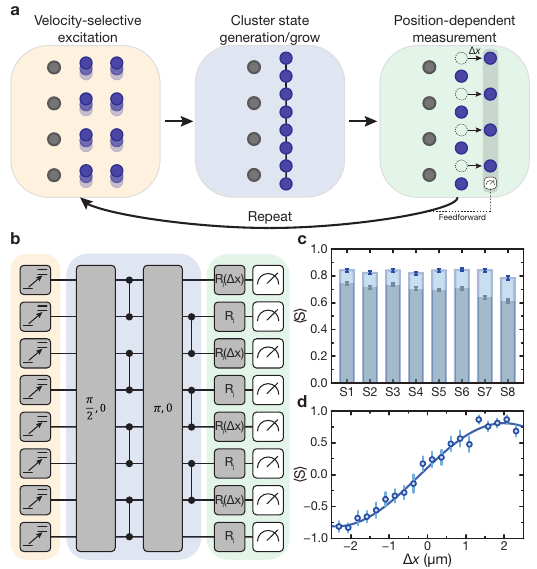}
\caption{\label{fig:4} \textbf{Cluster-state generation.} \textbf{a} Concept of using velocity selectivity and position-dependent phases for measurement-based quantum computing. Specific atoms (blue) out of a large reservoir array are moved towards an entangling zone and selectively excited into the qubit manifold. Entangling operations are used to generate a new cluster state or grow an existing one. Atom-specific measurement bases are realized according to the position of the atoms during a final single-qubit rotation. The atoms can then be selectively measured, re-cooled and recycled to repeat the process. \textbf{b} We experimentally demonstrate the basic building blocks of this quantum computing approach. Eight atoms are selectively excited into the qubit manifold with four stationary atoms remaining in the ground state (not shown). The eight atoms are entangled into a linear cluster state, with stabilizers $S_i = X_i \prod_{j \in \mathcal{N}(i)} Z_j$ which consist of a Pauli $X$ gate for each qubit and Pauli $Z$ gates for its neighbors $\mathcal{N}(i)$. \textbf{c} The stabilizers are measured with the help of local control through atom displacements. The local $R_j$ rotations consist of an echo sequence (Fig. \ref{fig:3}b), with atoms displaced during the $\pi$ pulse. We obtain average stabilizer expectation values of 0.830(4) and 0.694(4) with and without state-resolved detection, respectively. \textbf{d} Scanning the displacement of half the atoms, the measured expectation value oscillates, giving the stabilizer value at $\lambda_{FS}/8$ displacement.}
\end{figure}

The infidelity due to off-resonant (de-)excitation of stationary atoms is given as $\pi^2\,\mathrm{sinc}^2\left( \frac{\pi}{2} \sqrt{1 + (\Delta/\Omega)^2} \right)/4$ for a $\pi$-pulse acting on the moving atoms with Rabi frequency $\Omega$.
We can further relate the ratio of detuning to Rabi frequency with the total distance traveled during a $\pi$-pulse by the moving atoms, $d$, obtaining $d/\lambda_0=\Delta/(2\Omega)$.
The infidelity incurred on stationary atoms is therefore only a function of the distance traveled by the moving atoms during the pulse (Fig.~\ref{fig:2}d).
%
Move distances of a few microns are sufficient for high-fidelity operations, and fine-tuned detunings that correspond to $2\pi$-rotations on stationary atoms can be achieved with lower velocities (Fig.~\ref{fig:2}d, bottom) and sub-\SI{}{\micro\meter} moves.
The concept of velocity selectivity is thus general and can be applied to various platforms and qubit encodings (see Fig.~\ref{fig:2}d), with the narrow-line transitions in alkaline-earth and alkaline-earth-like atoms particularly well suited for local selective state preparation and readout. 
Velocity-selective state-preparation is also fully compatible with three-photon excitation from ${^1}\text{S}_0$ to ${^3}\text{P}_0$ under triple-magic conditions and at a significantly larger Rabi frequency of $\Omega=2\pi\times\SI{40}{\kilo \hertz}$ compared to the direct single-photon excitation~\cite{ammenwerth2025realization,he2025coherent,carman2025collinear}.
We demonstrate selective three-photon control by performing spectroscopy of the three-photon transition for atoms moving at a constant velocity of $v=\SI{0.052}{\meter\per\second}$ and stationary atoms nearby (Fig.~\ref{fig:2}e).
Here, the velocity is fine-tuned to minimize the impact on stationary atoms by driving a $2\pi$-rotation on them, while keeping the move velocity low.

Going beyond velocity-selective coherent control, we demonstrate a first step towards a velocity-selective qubit reset~\cite{lis2023midcircuit}, which is relevant for measurement-free quantum error correction approaches~\cite{heussen2024measurement,butt2025measurement}.
A dissipative transition from the qubit manifold to the ground state can be implemented via a two-photon transition to ${^3}\text{P}_1$, followed by a spontaneous decay~\cite{ammenwerth2025realization}.
We separately implement this transition for both qubit states. As the transition is dissipative, we increase the efficiency of population transfer by driving each transition with three pulses, while letting the atoms decay to the ground state in between during a wait time of \SI{40}{\micro\second} (Fig.~\ref{fig:2}f).
A mid-circuit measurement is then performed, confirming the transfer of both qubit states to the ground state.
A residual few-percent effect on stationary atoms is observed, mainly due to the dissipative nature of the transition.
As the moving atoms are intended to be cooled and reset, one could potentially move the atoms at higher velocities compared with the \SI{0.26}{\meter\per\second} used here, even at the expense of inducing heating, to reduce the undesired effect on stationary atoms.

Next, we demonstrate the capability of dynamical local control of single-qubit rotations.
While velocity selectivity requires a sizable momentum transfer as realized in the clock qubit, this approach is less effective for the co-propagating Raman beams that drive single-qubit rotations for the FS qubit.
We therefore take a different approach and use the spatially varying phase along the propagation direction of the Raman beams for implementing local $Z$ rotations.
This concept was recently demonstrated in the context of metrology for optical qubits using atom displacements with \SI{}{\nano\meter}-scale precision~\cite{shaw2024multi}.
For the FS qubit, with a qubit energy of \SI{17}{\tera\hertz}, the effective wavelength is $\lambda_{FS}\approx\SI{17.2}{\micro\meter}$.
Therefore, $\pi$-rotations can be achieved via convenient atom displacements on the order of a few microns, relaxing spatial precision and stability requirements compared with optical qubits.
We confirm the expected spatial phase pattern for local control experimentally by performing a Ramsey sequence where atoms trapped in AOD tweezers are displaced by a varying distance before the second $\pi/2$-pulse (Fig.~\ref{fig:3}a). 
Indeed, phase accumulation consistent with the expected FS wavelength is observed along the path of the Raman beams, with negligible phase shift in the perpendicular direction.

We demonstrate local rotations via displacement using an echo sequence simultaneously performed on three sets of atoms (Fig.~\ref{fig:3}b).
Four atoms are placed in stationary traps (green), while two additional sets of four atoms (blue and orange) are displaced by $\lambda_{FS}/8$ and $\lambda_{FS}/4$ before the $\pi$ echo pulse, respectively.
A position-dependent shift of the Ramsey signal is observed, confirming the ability to perform local $Z$ rotations.
The relatively high Rabi frequency of the single-qubit control beams ($\Omega_{FS}=2\pi\times\SI{120}{\kilo\hertz}$ in this measurement) together with the relatively long wavelength of the FS qubit further allow us to perform the locally controlled single-qubit rotations while the atoms are moving.
This can be understood from the fact that the spatial laser phase experienced by the moving atom does not significantly change during the gate, or equivalently, because the Doppler shift is small compared with the Rabi frequency.
We experimentally demonstrate the ability to perform local rotations on-the-fly by repeating the same echo sequence, with atoms stationary during the first two pulses, but with one set of atoms (blue) moving at a constant velocity of \SI{0.1}{\meter\per\second} during the final $\pi/2$-pulse (Fig.~\ref{fig:3}c).
Timing the move with respect to the final $\pi/2$-pulse results in a different phase due to the atom moving by an additional distance.
The observed Ramsey signal has a contrast of \SI{95.5(6)}{\percent}, indistinguishable from a reference measurement obtained with stationary atoms by scanning the phase of the second laser pulse (\SI{96.2(6)}{\percent}), showing the ability to have on-the-fly local control of the FS qubit.

The demonstrated velocity-selective state preparation and measurement in combination with local single-qubit rotations open the door to exploring two key applications for measurement- and circuit-based quantum computing.
In MBQC, a large computationally-useful resource state, such as the cluster state, is generated and the computation is performed through a series of single-qubit measurements and classical feedforward~\cite{briegel2009measurement}.
Importantly, the large cluster state required for the entire computation does not have to be generated at once, but can be extended during the computation, while other parts of the state have already been measured~\cite{schon2005sequential}.
The demonstrated tools are essential building blocks for the following velocity-enabled MBQC scheme (Fig.~\ref{fig:4}).
Consider a reservoir of atoms in the ground state, trapped either in optical tweezers as in our experiment, or potentially in a large-scale optical lattice which can be further continuously loaded~\cite{gyger2024continuous,norcia2024iterative,chiu2025continuous,li2025fast,muniz2025repeated}. In each step of the computation, selected atoms from the reservoir are picked up by movable tweezers and selectively excited into the qubit manifold while being moved to a nearby entangling zone.
Single and two-qubit gates are then performed in parallel to generate an initial cluster state, or grow an existing one.
Then, selected atoms from the cluster state are rotated to the desired measurement bases by placing them in different positions along the path of the control Raman beams during a pulse.
Finally, they are selectively transferred to the ground state to be measured.
With high-survival and fast imaging available~\cite{muzi2025microsecond}, information can be extracted rapidly and processed while measured atoms are cooled.
Subsequently, the cooled atoms can be used to refill the reservoir.
The process can be repeated with classical feed-forward operations in between until the entire computation is performed.
%

\begin{figure}[t]
\centering
\includegraphics[width=1\columnwidth]{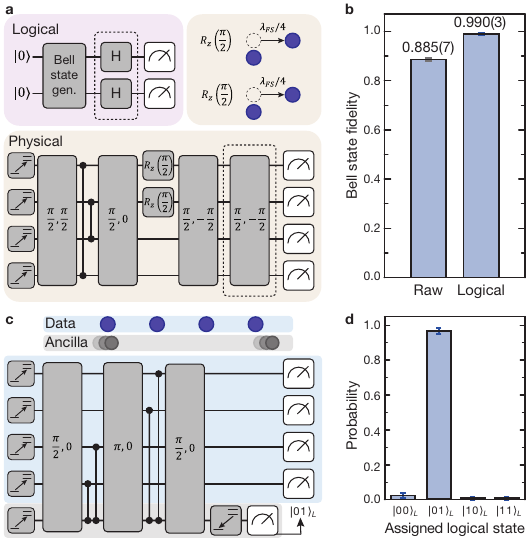}
\caption{\label{fig:5} \textbf{Entanglement between logical qubits.} \textbf{a} We realize a $[[4,2,2]]$ error-detection code and prepare a logical Bell state, whose fidelity is lower bounded using measurements in the $X$ (dashed box) and $Z$ bases. A fault-tolerant generation of the logical Bell state using four qubits is performed with the help of local control by atom displacement. A logical fidelity of \SI{99.0(3)}{\percent} is observed after post-selecting on the parity of the result, significantly exceeding the fidelity of individual physical Bell states prepared on pairs of physical qubits in the same sequence (\textbf{b}). \textbf{c} A flying ancilla moving at a constant velocity is used to prepare the logical $\ket{01}_L$ state of the $[[4,2,2]]$ code. The ancilla qubit is selectively excited into the qubit manifold, performs entangling operations with all data qubits using fly-by CZ gates, and is measured with a fast image following a selective three-photon transition into the ground state. Upon measurement of the ancilla qubit in the fast image, we identify the correct state of the $[[4,2,2]]$-code with a probability of \SI{96(2)}{\percent}.}
\end{figure}

We experimentally demonstrate the basic building blocks of this concept.
Twelve atoms acting as the initial reservoir are trapped in static optical tweezers.
Eight of them are picked up by movable tweezers and excited into the qubit manifold through a velocity-selective clock pulse, while the other four remain in the ground state. Half the atoms are then placed in static tweezers, while the other half remain in movable ones. Global single-qubit rotations and two-qubit gates are used to generate an eight-qubit linear cluster state (Fig.~\ref{fig:4}b).
The connectivity of the cluster state is set by the position of the movable tweezers with respect to the static ones during the two layers of parallel entangling gates.
Finally, stabilizer measurements are performed to benchmark the entanglement in the generated state. Measurement bases of different atoms are locally chosen using $Z$ rotations implemented via displacement of atoms along the propagation path of the Raman beams, using the echo pulse sequence shown in Fig.~\ref{fig:3}b.
All stabilizers, both with (blue) and without (gray) post-selecting on all atoms being present in the final state-resolved detection, are above $0.5$ (Fig.~\ref{fig:4}c), certifying the inseparability of the cluster chain~\cite{toth2005entanglement,bluvstein2022quantum,white2026quantum}.
State-resolved detection is particularly useful in the context of MBQC, as loss can be detected in every step and efficiently accounted for with suitable QEC codes~\cite{raussendorf2007topological,auger2018fault}.
%
%
Changing the displacement of half of the atoms used for the stabilizer measurement, the average expectation value of the measured correlators, each involving a single displaced atom, oscillates sinusoidally (Fig.~\ref{fig:4}d)~\cite{raussendorf2003measurement}.

As a second example, we explore the perspectives of the developed capabilities as key building blocks of circuit-based quantum computation and QEC with two instances of the $[[4,2,2]]$ error-detection code~\cite{vaidman1996error,reichardt2024fault,chung2025fault,zhang2025leveraging} corresponding to four logical qubits in total.
We first generate a logical Bell-state encoded within each $[[4,2,2]]$ code block using a fault-tolerant circuit, with local rotations implemented by moving specific atoms along the path of the Raman beams~\cite{chung2025fault} (Fig.~\ref{fig:5}a).
We perform measurements of all qubits in either the $Z$ or $X$ bases (using a transversal Hadamard gate~\cite{chung2025fault}) to obtain a lower bound for the fidelity of the logical state $F_L\ge(\avg{X_LX_L} + \avg{Z_LZ_L})/2$~\cite{guhne2009entanglement}.
The parity of the physical measured state in each basis is used to detect errors in the $[[4,2,2]]$ code.
With only positive-parity data that is compatible with the stabilizers of the code taken into account (\SI{10.6}{\percent} of the data discarded), we obtain a logical fidelity of \SI{99.0(3)}{\percent} (Fig.~\ref{fig:5}b), significantly higher than the raw Bell-state fidelity of physical pairs prepared in the same sequence (\SI{88.5(7)}{\percent}). In addition, the logical Bell fidelity is also slightly higher than our best fidelity achieved in the same platform on physical qubits~\cite{tao2025universal}.

Finally, we combine single-qubit on-the-fly control with fly-by CZ gates to realize stabilizer measurements with a flying ancilla.
The four atoms encoding the $[[4,2,2]]$ code, as well as an ancilla atom, are selectively excited using a clock pulse.
To generate the $\ket{01}_L$ logical state of the $[[4,2,2]]$ code, the ancilla atom is moved at a constant velocity of $v=\SI{0.1}{\meter\per\second}$ along the chain of data qubits, at a distance of \SI{2}{\micro\meter} away from them.
The ancilla atom is entangled with all data qubits using fly-by CZ gates, and then its ${^3}\text{P}_0$ population is measured using a velocity-selective three-photon transition to the ground state followed by a fast mid-circuit detection (Fig.~\ref{fig:5}c).
Then, a state-resolved measurement of the data qubits is performed, and the generated logical state is assigned according to the physical state of the four qubits (Fig.~\ref{fig:5}d).
Heralded by a successful mid-circuit measurement of the ancilla qubit, the correct logical state is measured in \SI{96(2)}{\percent} of valid outcomes (\SI{14}{\percent} of shots are discarded due to wrong parity). More information about fly-by CZ gates and results for repeated ancilla measurements are provided in the SI~\cite{SI}.

%
%
In conclusion, we have demonstrated a novel velocity-enabled architecture for fast quantum computing with neutral atoms.
While our experimental demonstration is based on the FS qubit in strontium, the presented ideas are general and could be translated to any neutral-atom platform with suitable adjustments.
For instance, in hyperfine qubits where the effective wavelength is on the millimeter scale~\cite{bluvstein2022quantum}, one can change the relative angle between the control Raman beams to set a desired effective wavelength along a given propagation axis that will allow for local control via displacement or velocity-selectivity.
Here, however, the absence of a decoupled stable state complicates site-selective and non-destructive imaging, which might require a dedicated spatial zone with long-distance shuttling and hiding beams~\cite{chiu2025continuous}, possibly combined with cavity-based readout~\cite{deist2022mid,grinkemeyer2025error,hu2025site}.
In addition, while our approach allows for a spatial-zone-free architecture with the use of only global control beams aside from the optical tweezers, focused, potentially dynamically controlled~\cite{rines2025demonstration} beams as well as several spatial zones~\cite{bluvstein2022quantum,bluvstein2025fault} can be used together with the ideas presented here for increased flexibility.
For instance, velocity selectivity as well as local single-qubit control can be realized within each zone to reduce the time-overhead of shuttling atoms between zones, while dedicated zones for entangling gates or readout can still be used to reduce the required Rydberg or imaging laser powers.

Beyond our current demonstration, on-the-fly operations on atoms moving in curved trajectories as well as SWAP operations transferring information from moving atoms to stationary or counter-propagating ones without stopping can also be envisioned, opening the path to repeated syndrome readout with minimal ancilla qubit overhead.
In the future, fast and efficient atom shuttling~\cite{Lam2021Demonstration,Hwang2025Fast,Kim2025Blinking} as well as hardware improvements such as using fast spatial light modulators~\cite{ammenwerth2025dynamical, lin2025ai,wei202610,bytyqi2026device} for simultaneously moving multiple atoms in different velocities and directions can further increase the scalability of our approach beyond standard AODs.
In the compilation layer, velocity-enabled computing opens a new dimension for optimization at the physical or logical circuit level, for example to find optimal QEC codes that maximally leverage the platform's benefits, or applications that can be run efficiently on the platform.
Finally, we anticipate that a platform based on velocity-enabled fly-by operations with the qubits stored in moving optical lattices or large-scale tweezer arrays can overcome current scaling limits in neutral-atom quantum computers and pave the way to scalable quantum computers based on neutral atoms.
%

\begin{acknowledgments}
		We acknowledge funding by the Max Planck Society (MPG), the Deutsche Forschungsgemeinschaft (DFG, German Research Foundation) under Germany's Excellence Strategy--EXC-2111--390814868, and through JST-DFG2024: Japanese-German Joint Call for Proposals on “Quantum Technologies” (Japan-JST-DFG- ASPIRE 2024) under DFG Grant No. 554561799, from the Munich Quantum Valley initiative as part of the High-Tech Agenda Plus of the Bavarian State Government, and from the BMFTR through the programs MUNIQC-Atoms and MAQCS.
		This publication has also received funding under Horizon Europe programme HORIZON-CL4-2022-QUANTUM-02-SGA via the project 101113690 (PASQuanS2.1).
		J.Z. acknowledges support from the BMFTR through the program “Quantum technologies---from basic research to market” (SNAQC, Grant No. 13N16265).

        O.L. acknowledges support from the Rothschild and CHE Quantum Science and Technology fellowships. H.T., M.A. and R.T. acknowledge funding from the International Max Planck Research School (IMPRS) for Quantum Science and Technology. M.A. acknowledges support through a fellowship from the Hector Fellow Academy.
		F.G. acknowledges funding from the Swiss National Fonds (Fund Nr. P500PT\underline{\hspace{2mm}}203162).
        Competing interests: J.Z. is a co-founder and shareholder of PlanQC GmbH.

	\end{acknowledgments}


\bibliography{bibliography}
\clearpage

\clearpage
\appendix
\section*{Supplementary Information}

\setcounter{figure}{0}
\renewcommand{\thefigure}{S\arabic{figure}}

\renewcommand{\theHfigure}{S\arabic{figure}}

\setcounter{equation}{0}
\renewcommand{\theequation}{S\arabic{equation}}

\renewcommand{\theHequation}{S\arabic{equation}}

\section{Details about the experimental apparatus}

Our experimental apparatus traps single strontium atoms in static and dynamic optical tweezers at a wavelength of \SI{813}{\nano\meter} (Fig.~\ref{fig:S1}a), which is a magic wavelength for the ${^1}\text{S}_0$ to ${^3}\text{P}_0$ clock transition. Sisyphus cooling on the intercombination line enables high-fidelity, low-loss imaging~\cite{tao2025universal}. Sideband cooling is further used to bring atoms close to their radial motional ground state ($\bar n\approx0.1$). Fast imaging is performed destructively within a few tens of \SI{}{\micro\second}, without cooling, using counter-propagating pulses at \SI{461}{\nano\meter}. A \SI{496}{\nano\meter} repumper is used to selectively transfer atoms from ${^3}\text{P}_2$ to the ground state, with minimal coupling to the ${^3}\text{P}_0$ qubit state, enabling state-resolved detection (SRD).

State preparation in the ${^3}\text{P}_0$ state is achieved either via a single-photon transition using a \SI{10}{\milli\watt} clock laser at \SI{698}{\nano\meter} and a magnetic field of \SI{420}{\gauss}, or a three-photon transition as described below. For coherent operations, the fine-structure (FS) qubit is trapped in magic conditions by tuning a \SI{19}{\gauss} magnetic field to an in-plane angle of \ang{79} with respect to the tweezer polarization~\cite{ammenwerth2025realization}.

More information about the above experimental procedures can be found in our recent publications~\cite{Tao2024,ammenwerth2025realization,rines2025demonstration}. In this work, we further introduce several experimental upgrades. We align the polarization of movable tweezers with that of static tweezers, to allow for simultaneous coherent trapping of the FS qubit in magic conditions in both static and dynamic traps. We demonstrate and benchmark coherent shuttling of the FS qubit in an echo Ramsey sequence, showing that the contrast of the Ramsey signal is preserved for tens of moves with a length of \SI{17.2}{\micro\meter}. More specifically, we move back and forth, with a $\pi$-pulse performed at the return point (Fig.~\ref{fig:S1}b). A single move is defined as a single stroke between two such return points. In addition, we improve our CZ gate fidelity, and make significant changes to our Raman and three-photon laser systems to enable higher Rabi frequencies, as described in the next sections.

\subsection{Raman and three-photon beams}

\begin{figure*}[t!]
\centering
\includegraphics[width=1\textwidth]{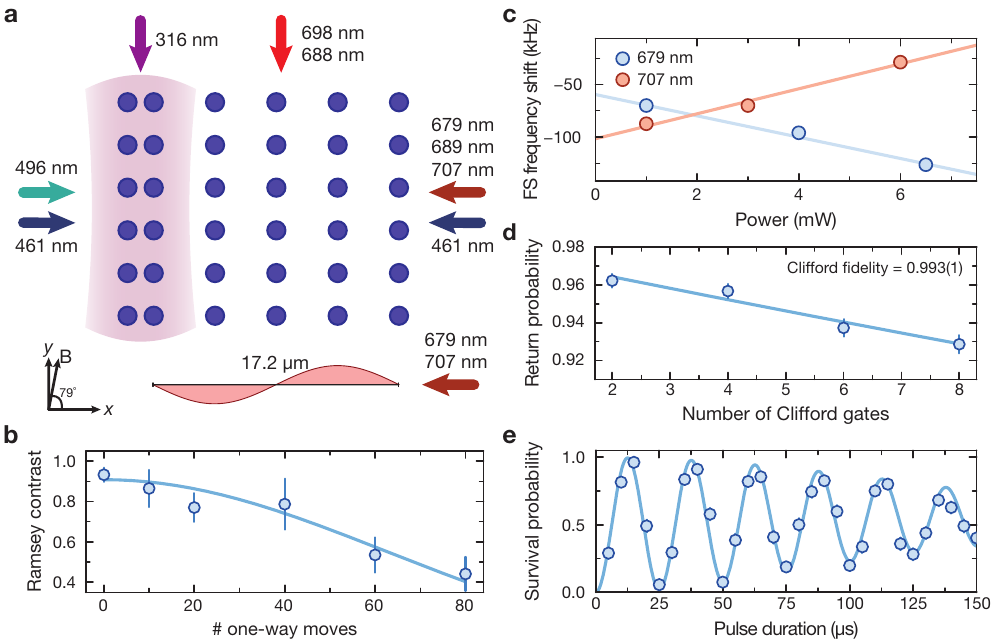}
\caption{\label{fig:S1} \textbf{Experimental details.} \textbf{a} A sketch of the experimental setup. The \SI{813}{\nano\meter} static trap sites (blue circles) are marked (not to scale). The two leftmost columns, which are used for calibrating and benchmarking CZ gates in static traps, are \SI{2}{\micro\meter} apart in the horizontal direction and \SI{13}{\micro\meter} apart vertically. The rest of the columns share the same vertical spacing, but are horizontally spaced by \SI{4.3}{\micro\meter}, which is a quarter of the FS wavelength. These columns are currently only used as a reservoir to sort defect-free arrays in the leftmost columns, or to store atoms that will then be transferred to dynamic tweezers used in the computation. All measurements involving CZ gates are thus performed in the entangling zone, with static atoms stored in the second column from the left, and dynamic atoms moving nearby. Directions of relevant beams as well as the orientation of the magnetic field with respect with the tweezer polarization ($x$-direction) are presented. \textbf{b} Ramsey contrast as a function of one-way \SI{17.2}{\micro\meter}-long moves of the FS qubit with a $\pi$-pulse performed at the return point. A fit for Gaussian decay is plotted to guide the eye, with coherence being preserved for tens of moves ($90(10)$ fitted). \textbf{c} Detuning of the Raman FS transition as a function of power of both the \SI{679}{\nano\meter} and \SI{707}{\nano\meter} beams is presented, where the power of the other beam remains fixed. A magic power ratio is found according to the slopes of the linear fits. \textbf{d} Clifford randomized benchmarking of single-qubit gates at a Rabi frequency of \SI{200}{\kilo\hertz}. \textbf{e} Rabi oscillations for the three-photon transition between the ground state and ${^3}\text{P}_0$, at \SI{600}{\micro\kelvin} traps, with ground state atoms pushed after the evolution. The number of coherent Rabi oscillations is currently limited by scattering from the ${^3}\text{P}_1$ state.
}
\end{figure*}

We perform single-qubit gates on the FS qubit using co-propagating \SI{679}{\nano\meter} and \SI{707}{\nano\meter} lasers, detuned by \SI{12}{\giga\hertz} from the ${^3}\text{S}_1$ state. As both lasers induce light shifts with opposite signs on the FS qubit, we choose a magic power ratio where the effective shift in qubit energy vanishes (Fig.~\ref{fig:S1}c). In contrast to our previous work, we use External-cavity Diode Lasers (ECDLs) locked to the same high-finesse cavity instead of a frequency comb. This allowed us to push the Pound–Drever–Hall servo-bump above a \SI{}{\mega\hertz}, reducing the phase noise in the relevant frequency range between $100\,$kHz and $1\,$MHz.

Using injection-locked amplifiers for each beam, we could therefore increase our previously used Rabi frequency of \SI{20}{\kilo\hertz} by up to two orders of magnitude, working in the frozen-atom regime. In Fig.~\ref{fig:S1}d, we show randomized benchmarking of the single-qubit Clifford gate fidelity at a Rabi frequency of \SI{200}{\kilo\hertz}, achieving a comparable fidelity to our previous work~\cite{tao2025universal} with an order-of-magnitude faster gates. For the experimental results shown in this work, we use approximately \SI{4}{\milli\watt} of power per beam, resulting in a Rabi frequency of \SI{120}{\kilo\hertz} for an elliptical beam size of \SI{130}{\micro\meter} by \SI{1.2}{\milli\meter} in the atomic plane. We expect higher-fidelity gates with a \SI{}{\mega\hertz}-scale Rabi frequency to be achievable in the near future by utilizing feed-forward for phase-noise suppression~\cite{chao2025robust}.

For state-preparation and measurement, a three-photon transition coupling the ground state and ${^3}\text{P}_0$ is used~\cite{ammenwerth2025realization,he2025coherent,carman2025collinear}. To this end, ECDLs at \SI{688}{\nano\meter} and \SI{689}{\nano\meter} are locked to different bores of the same high-finesse cavity to which the FS lasers are locked. The \SI{688}{\nano\meter} laser is amplified using an injection-locked amplifier. We use a red detuning of \SI{12}{\giga\hertz} from ${^3}\text{S}_1$ and \SI{20}{\mega\hertz} from ${^3}\text{P}_{1,m_J=-1}$. The powers of the \SI{679}{\nano\meter}, \SI{688}{\nano\meter}, and \SI{689}{\nano\meter} lasers are \SI{4.5}{\milli\watt}, \SI{10}{\milli\watt}, and \SI{1}{\milli\watt}, respectively. The \SI{689}{\nano\meter} beam is co-propagating with the FS beams, albeit at an orthogonal polarization. The \SI{688}{\nano\meter} beam is launched from an orthogonal direction, having a beam diameter of \SI{100}{\micro\meter}. Directions of relevant laser beams used in the experiment are shown in Fig.~\ref{fig:S1}a. We obtain a state-preparation fidelity of \SI{97}{\percent} with a Rabi frequency of approximately \SI{40}{\kilo\hertz} (Fig.~\ref{fig:S1}e).

Two-photon transitions to ${^3}\text{P}_1$ are performed using the same laser setup, beam directions, and detuning from ${^3}\text{S}_1$. In this case, the \SI{688}{\nano\meter} power is reduced by over two orders of magnitude to avoid scattering of atoms in ${^3}\text{P}_1$ through ${^3}\text{S}_1$, while \SI{679}{\nano\meter} and \SI{707}{\nano\meter} powers remain the same as for controlling the FS qubit.

\begin{figure*}[t!]
\centering
\includegraphics[width=1\textwidth]{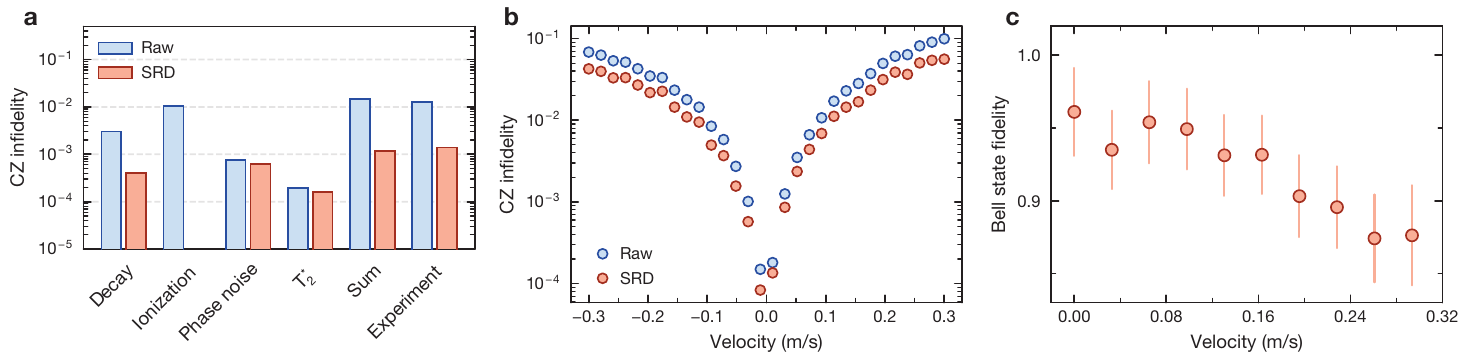}
\caption{\label{fig:S2} \textbf{Static and fly-by CZ gates.} \textbf{a} Error budget for static CZ gates at \SI{5}{\mega\hertz} Rabi frequency, with and without state-resolved detection (SRD). \textbf{b} Infidelity due to the Doppler effect for a fly-by CZ gate performed with the parameters of a stationary gate. \textbf{c} Experimentally measured Bell-state fidelity as a function of atom velocity during the fly-by CZ gates. The Bell-state is prepared by applying a $\pi$-half gate, fly-by CZ gate, and a final $\pi$-quarter gate. The fidelity is estimated by measuring the populations and parity oscillation for each velocity.}
\end{figure*}

\section{Static and fly-by CZ gates}

We perform our CZ gates via a time-optimal pulse coupling the ${^3}\text{P}_0$ qubit state with the $n=47\,{}^3\text{S}_1, m_J = -1$ Rydberg state. The UV laser system used for driving the Rydberg transition is identical to our previous demonstration~\cite{tao2025universal}. However, we empirically observe that humidity fluctuations in our lab affect the overall stability of the gate performance. With an improved humidity control, a humidity stability on the order of \SI{2}{\percent} is achieved. We observe that under these conditions, the beam position and Rydberg resonance are stable enough throughout the day to allow for high-fidelity gates. 

Indeed, we observe that the optimal gate parameters are stable for over a day. The effective parameters found using the Hessian~\cite{muniz2025high,tao2025universal} are calibrated by maximizing the return probability in an echo sequence~\cite{tsai2025benchmarking} with ten CZ gates. The error budget at our \SI{5}{\mega\hertz} Rabi frequency is presented in Fig.~\ref{fig:S2}a and agrees well with the experimental results. Coupling the other qubit state, ${^3}\text{P}_2$, to the Rydberg state would eliminate the ionization of this state which limits our current fidelity before state-resolved detection~\cite{tao2025universal}. 
We expect that, even without loss correction, slightly higher Rabi frequencies and Rydberg states with higher $n$ will push gate fidelities beyond the \SI{99.9}{\percent} mark.

For fly-by CZ gates, additional sources of infidelity should be considered. As the distance traveled by the moving atom is limited to tens of nanometers during the fast gate, most potential sources of infidelity, such as Blockade change during the gate, become negligible for correct pulse timing. Since our UV beam and the \SI{688}{\nano\meter} beam used for velocity-selectivity in three-photon state-preparation and readout are co-propagating in our experiment, we focus on fly-by CZ gates with flying ancillas moving in the direction of the UV beam. The main infidelity in our case thus comes from the Doppler shift experienced by the moving atom with respect to its stationary counterpart (Fig.~\ref{fig:S2}b). We experimentally observe this effect by measuring the Bell-state fidelity obtained with fly-by CZ gates at different velocities (Fig.~\ref{fig:S2}c), and choose a velocity of $v=\SI{0.1}{\meter\per\second}$ for flying ancilla measurements where the infidelity is moderate. While beyond the scope of our demonstration, we expect fly-by gates using atoms moving perpendicularly to the UV beam to have significantly higher fidelities at large velocities.

\section{Syndrome measurement with a flying ancilla}

In the main text, we discussed the generation of the $\ket{01}_L$ state in a $[[4,2,2]]$ error-detection code using a flying ancilla. In that measurement, both the four data qubits and the flying ancilla are prepared using a velocity-selective clock pulse. The ancilla then performs four fly-by CZ gates with the data qubits, and is selectively measured by a fast image after mapping its ${^3}\text{P}_0$ population to the ground state with a velocity-selective three-photon pulse.

\begin{figure*}[t!]
\centering
\includegraphics[width=1\textwidth]{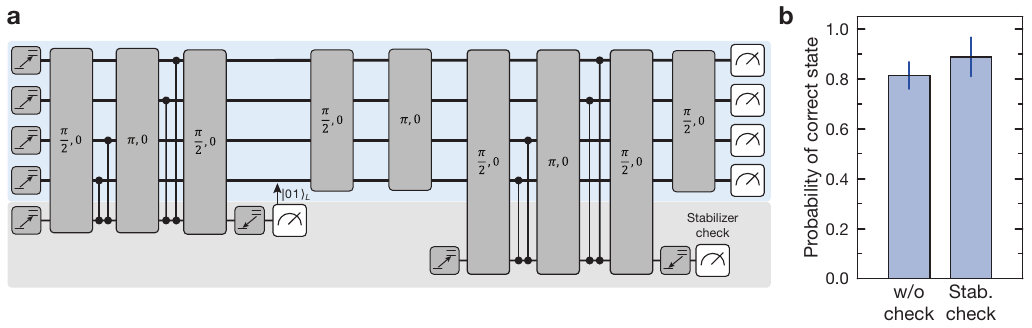}
\caption{\label{fig:S3} \textbf{Fly-by syndrome measurement.} \textbf{a} Circuit diagram for the generation and $ZZZZ$-syndrome measurement of the $\ket{01}_L$ logical state using two fly-by ancillas. \textbf{b} Probability of identifying the correct logical state with or without post-selecting on the syndrome measurement ancilla result.}
\end{figure*}

Here, we show additional data where, following the state preparation, the $ZZZZ$-stabilizer of the state is measured mid-circuit with a second flying ancilla (Fig.~\ref{fig:S3}a). As the qubits are now already under FS triple-magic conditions, the second ancilla is initialized into the ${^3}\text{P}_0$ state using a velocity-selective three-photon pulse. Global single-qubit gates and four fly-by CZ gates followed by a velocity-selective three-photon readout complete the stabilizer measurement. In this measurement, both ancillas are measured in the same mid-circuit fast image after being transferred to static tweezers. However, we note that measurement in separate images as well as in the AOD tweezers directly is possible. Potentially, fly-by swapping of information from the moving ancilla to a static readout qubit can also be envisioned. With suitable \SI{}{\micro\second}-scale readout duration (recently demonstrated for ytterbium~\cite{muzi2025microsecond}) and an appropriate imaging system, trapped ancillas could potentially be measured directly while moving sub-micron distances during an imaging pulse.

The correct logical state $\ket{01}_L$ is identified with a better fidelity when post-selecting on a correct stabilizer value in the mid-circuit syndrome measurement (Fig.~\ref{fig:S3}b). The overall fidelity is, however, lower compared with the measurement without the syndrome measurement block, as shown in the main text. The main source of infidelity in our case is the three-photon transitions, which limit the fidelity of the syndrome measurement block. Our three-photon $\pi$-pulse fidelity in deep traps is approximately \SI{97}{\percent}. However, the FS-qubit coherence is limited in deep traps~\cite{ammenwerth2025realization}, and therefore we typically use shallow \SI{30}{\micro\kelvin} traps for coherent FS manipulations. At such shallow traps, the three-photon Rabi frequency is similar to the trap frequency, and we observe significantly reduced fidelity. Therefore, we currently balance a tradeoff between coherence of the FS qubit and fidelity of mid-circuit state-preparation and readout when they are performed at the same trap depth. This technical issue will be resolved in future upgrades at higher Rabi frequency of the three-photon transfer, or alternatively using single-photon clock operations at a high-magnetic field under triple-magic conditions.

\section{Atom transfer between velocity zones}

Our approach of using velocity as a new degree of freedom for quantum computing offers a powerful extension of the standard spatial zone-based architecture. Spatially separated zones between which atoms are shuttled using their velocity are replaced with velocity zones between which atoms are accelerated. These changes, together with the demonstrated on-the-fly single-qubit rotations and fly-by CZ gates, provide several advantages.

On the hardware level, different velocity zones can be conveniently addressed by overlapping global beams at different detunings. In this case, velocity separation between zones determines the crosstalk, which is independent of the beam's shape and the spatial configuration of the atoms. In addition, velocity selectivity offers a potential reduction in space-time overhead during computation. While we leave the development of efficient compilation algorithms for our velocity-enabled approach for future work, one can already appreciate the potential time and space saving by considering the simple example of atom shuttling between zones.

In a standard spatial zone architecture, zones are typically separated by a scale of hundred microns, with shuttling times between different zones exceeding hundreds of microseconds~\cite{bluvstein2022quantum,manetsch2025tweezer}. For velocity zones, the equivalent metrics are the distance and time it takes to accelerate the atoms to the velocity of a different zone. As an example, we consider shuttling of atoms between spatial zones separated by \SI{100}{\micro\meter} with a \SI{200}{\micro\second}-long cubic trajectory. For velocity zones, assuming a Rabi frequency of \SI{40}{\kilo\hertz} for state-preparation and measurement, a velocity difference of approximately \SI{0.05}{\meter\per\second} between velocity zones is adequate. Using the same jerk as in the case of spatial zones, the acceleration of atoms between velocity zones takes approximately \SI{25.8}{\micro\second} over a moving distance of \SI{860}{\nano\meter}, saving an order of magnitude in time and two orders of magnitudes in shuttling distance. We expect these time and space savings to translate to applications in quantum computing and QEC using tailored compilers which will enable a more complete comparison between the different architectures.

\end{document}